\documentclass[twoside]{dis09}
\usepackage[latin1]{inputenc}
\usepackage[dvips]{graphicx,epsfig,color}
\usepackage{wrapfig,rotating}
\usepackage{amssymb,amsmath,array}

\begin{document}
\title{Open charm hadroproduction and the intrinsic charm content of the
proton}

\author{Bernd A. Kniehl
%
%
\vspace{.3cm}\\
%
II. Institut f\"ur Theoretische Physik, Universit\"at Hamburg,\\
Luruper Chaussee 149, 22761 Hamburg, Germany
%
\vspace{.1cm}\\
}

\maketitle

\begin{abstract}
We advocate charmed-hadron inclusive hadroproduction as a laboratory to probe
intrinsic charm (IC) inside the colliding hadrons.
Working at next-to-leading order in the general-mass variable-flavor-number
scheme endowed with non-perturbative fragmentation functions recently
extracted from a global fit to $e^+e^-$ annihilation data from KEKB, CESR, and
LEP1, we first assess the sensitivity of Tevatron data of $D^0$, $D^+$, and
$D^{*+}$ inclusive production to the IC parameterizations provided by Pumplin
{\it et al.}
We then argue that similar data from $pp$ collisions at RHIC would have the
potential to discriminate between different IC models provided they reach out
to sufficiently large values of transverse momentum.
\end{abstract}

\section{Introduction}

The general-mass variable-flavor-number scheme (GM-VFNS) provides a rigorous
theoretical framework for the theoretical description of the inclusive
production of single heavy-flavored hadrons, combining the
fixed-flavor-number scheme (FFNS) and zero-mass variable-flavor-number scheme
(ZM-FVNS), which are valid in complementary kinematic regions, in a unified
approach that enjoys the virtues of both schemes and, at the same time, is
bare of their flaws.
Specifically, it resums large logarithms by the
Dokshitzer-Gribov-Lipatov-Altarelli-Parisi (DGLAP) evolution of
non-perturbative fragmentation functions (FFs), guarantees the universality
of the latter as in the ZM-VFNS, and simultaneously retains the mass-dependent
terms of the FFNS without additional theoretical assumptions.
It was elaborated at next-to-leading order (NLO) for $e^+e^-$ annihilation
\cite{Kneesch:2007ey} photo- \cite{Kramer:2001gd} and hadroproduction
\cite{Kniehl:2004fy,Kniehl:2005st}.
In this presentation \cite{url}, we report recent progress in the theoretical
description of $D$-meson hadroproduction and argue that this process provides
a useful probe of IC in the proton \cite{Kniehl:2009ar}.
In Sec.~\ref{sec:ii}, we present updated predictions for the
transverse-momentum ($p_T$) distributions of $D^0$, $D^+$, and $D^{*+}$ mesons
\cite{Kniehl:2009ar} using our new FF sets \cite{Kneesch:2007ey}.
In Sec.~\ref{sec:iii}, we briefly review the constraints on IC obtained through
an extension \cite{Pumplin:2007wg} of the CTEQ6.5 global analysis
\cite{Tung:2006tb} endowed with three representative IC models.
In Sec.~\ref{sec:iv}, we investigate, for $D^0$ hadroproduction at the
Tevatron and at RHIC, the shifts in cross section produced by these IC
implementations.
Our conclusions are summarized in Sec.~\ref{sec:v}.

\boldmath
\section{$D$-meson hadroproduction revisited}
\label{sec:ii}
\unboldmath

\begin{figure}[ht]
\begin{center}
\begin{tabular}{ccc}
\parbox{0.3\textwidth}{
\epsfig{file=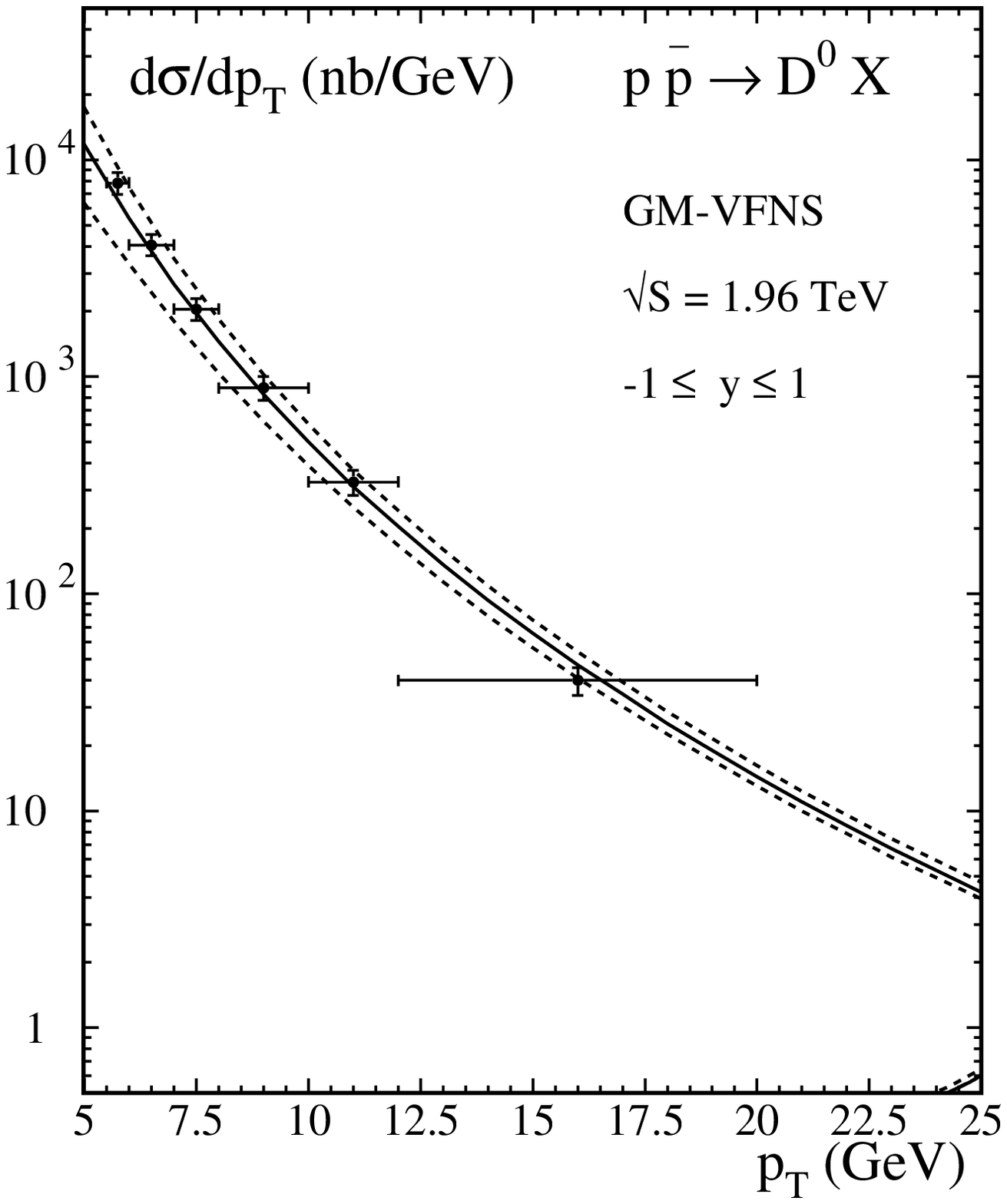,width=0.3\textwidth}}
&
\parbox{0.3\textwidth}{
\epsfig{file=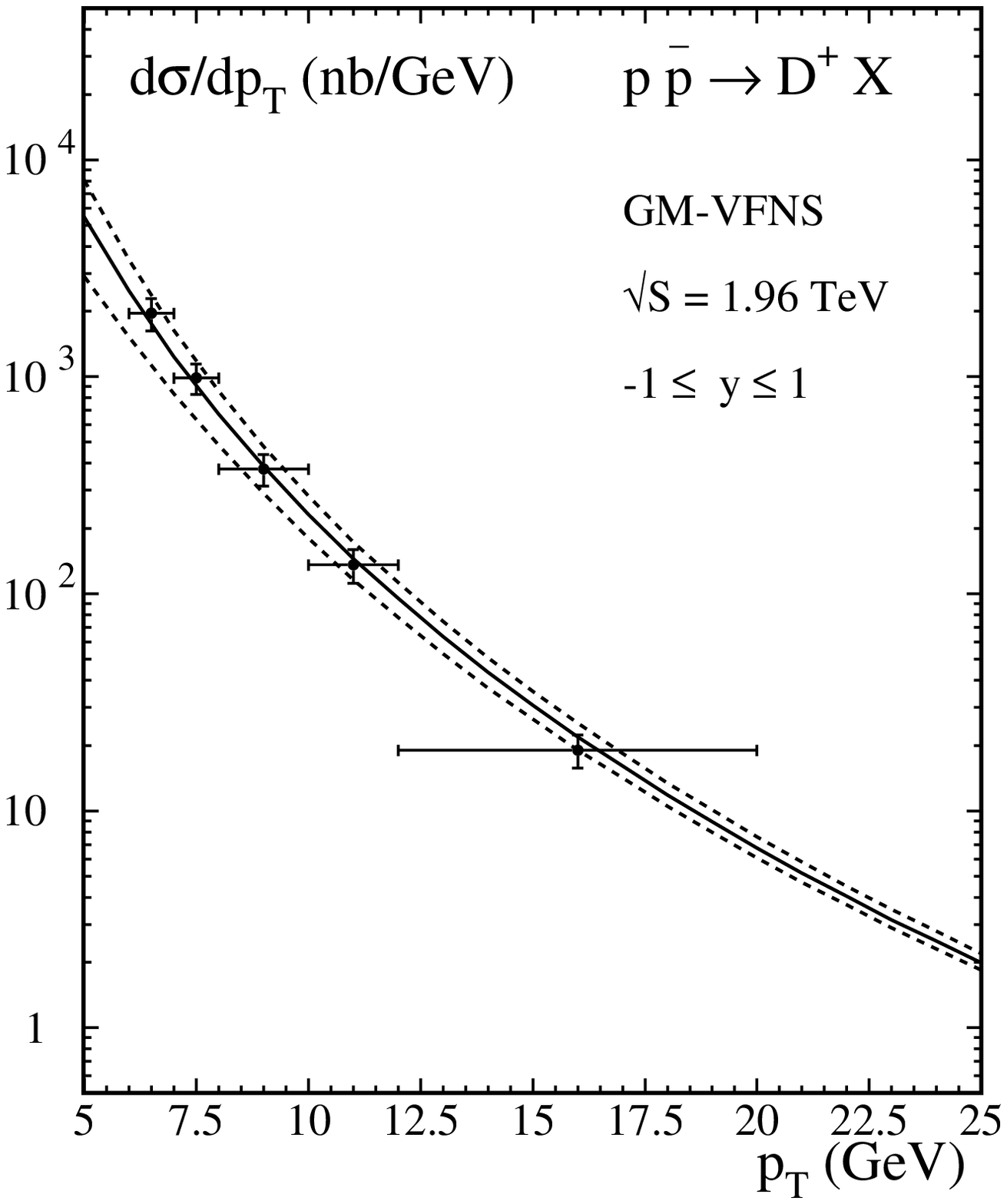,width=0.3\textwidth}}
&
\parbox{0.3\textwidth}{
\epsfig{file=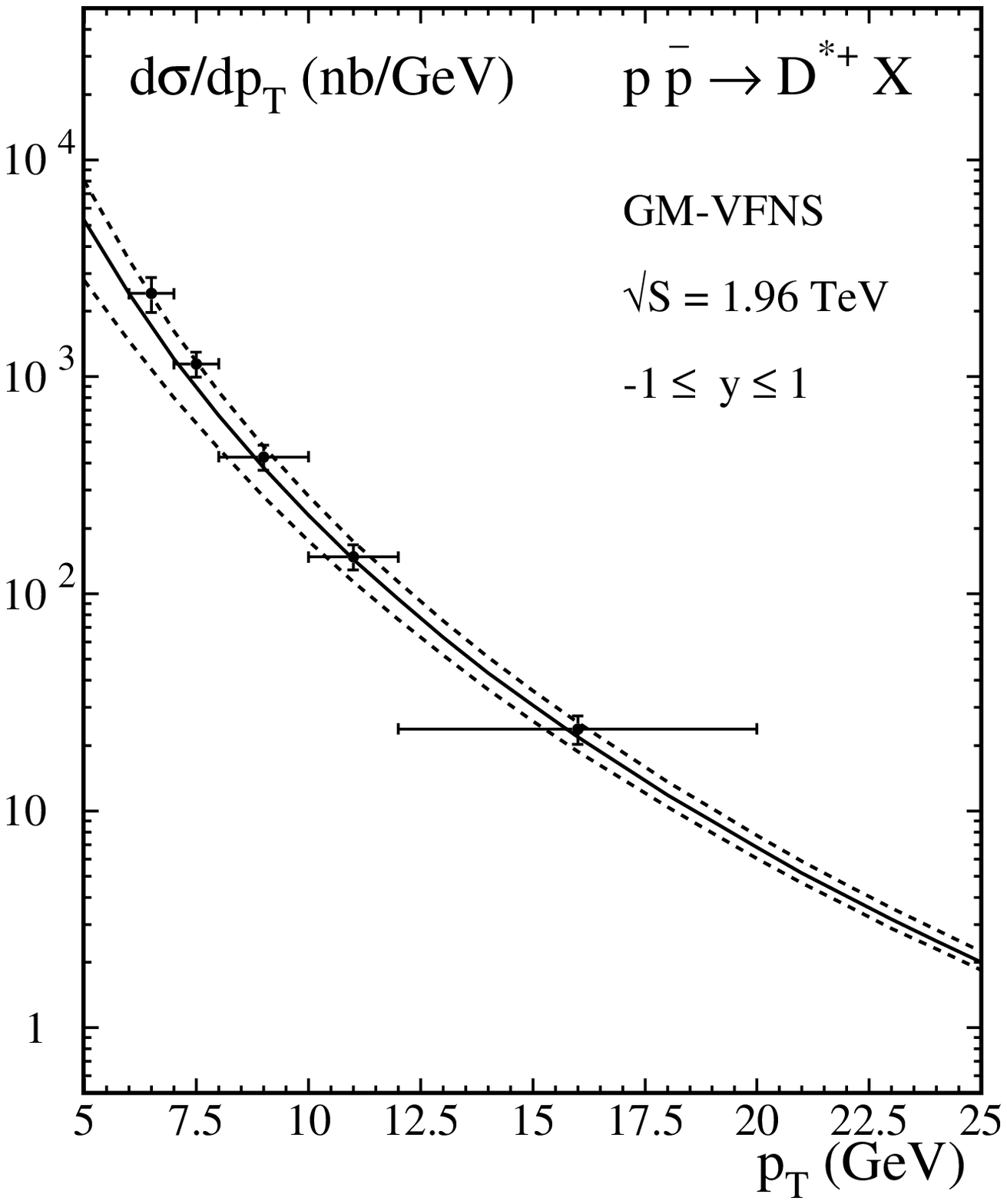,width=0.3\textwidth}}
\\
(a) & (b) & (c) \\
\\
\parbox{0.3\textwidth}{
\epsfig{file=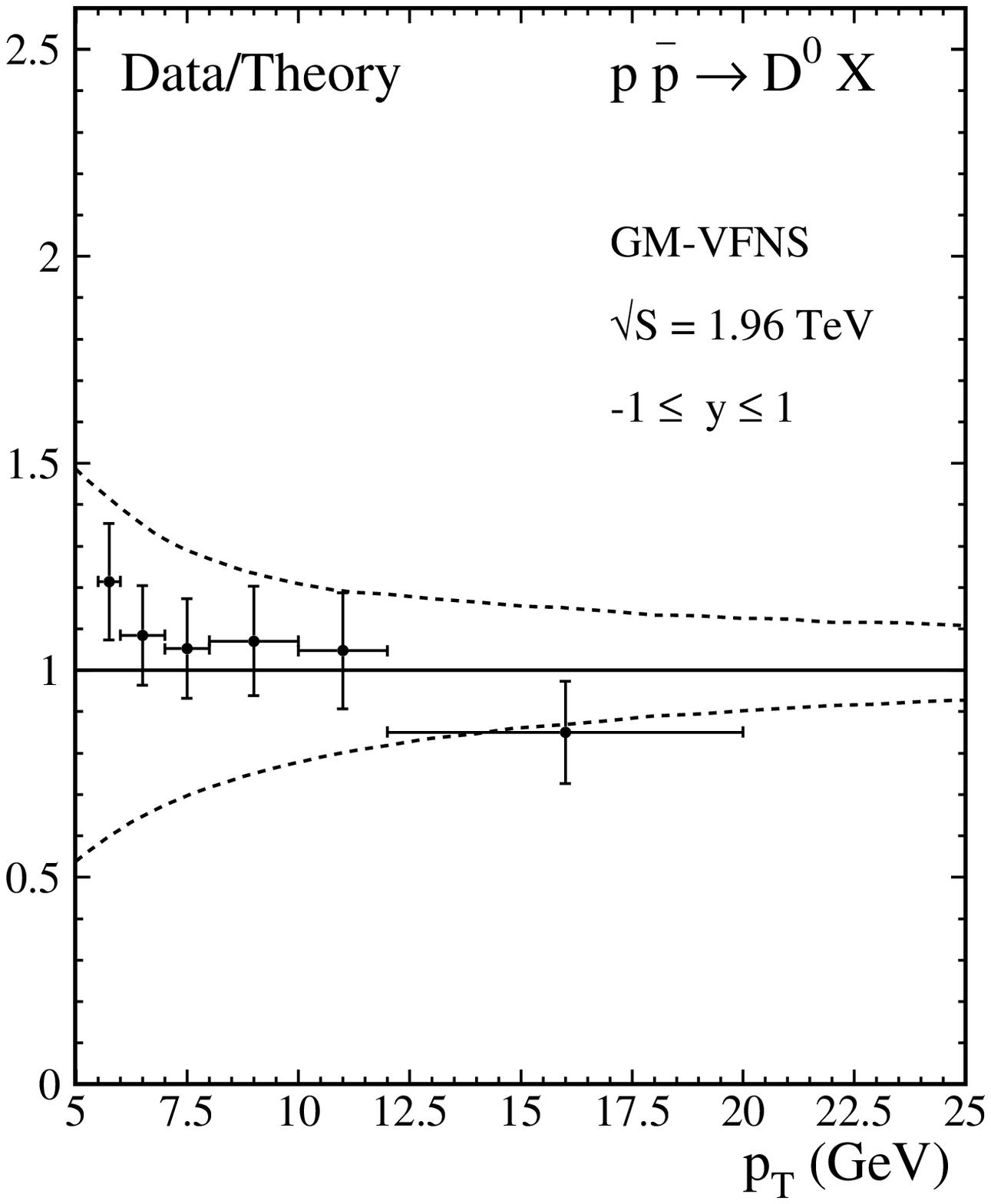,width=0.3\textwidth}}
&
\parbox{0.3\textwidth}{
\epsfig{file=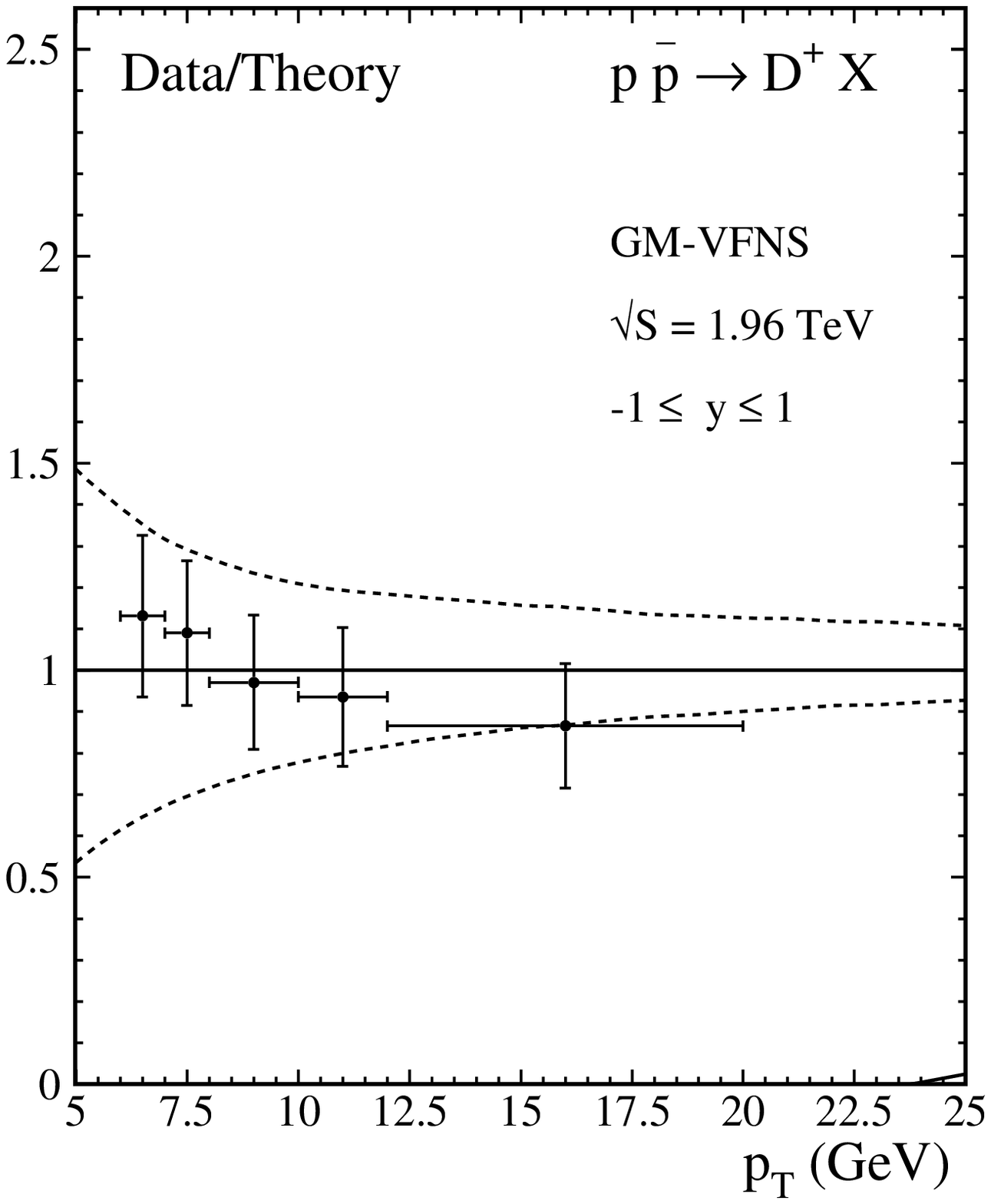,width=0.3\textwidth}}
&
\parbox{0.3\textwidth}{
\epsfig{file=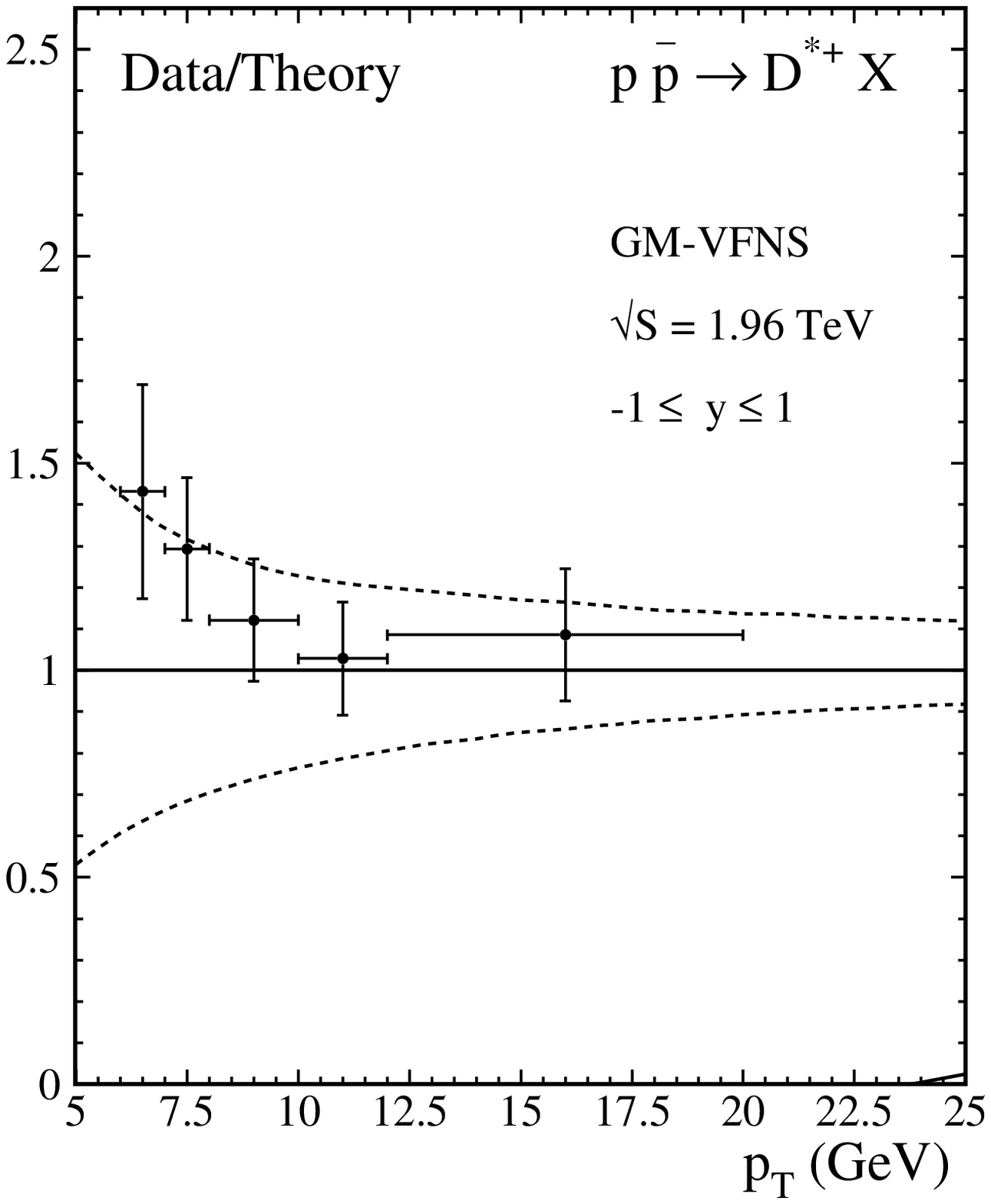,width=0.3\textwidth}}
\\
(d) & (e) & (f)
\end{tabular}
\end{center}
\caption{\label{fig:i}%
The solid lines represent the default predictions, and the dashed lines
indicate the theoretical uncertainty.}
\end{figure}

Our previous NLO predictions for inclusive $D$-meson hadroproduction
\cite{Kniehl:2005st} were evaluated with the FF sets of
Ref.~\cite{Kniehl:2006mw}.
These and previous FF sets \cite{Binnewies:1997xq} had been determined through
fits to LEP1 data \cite{Alexander:1996wy} in the ZM-VFNS.
These determinations have recently been updated and improved
\cite{Kneesch:2007ey} by also including KEKB and CESR \cite{Artuso:2004pj}
data, switching to the GM-VFNS, and also incorporating hadron-mass
corrections.
They greatly improve the agreement with the $p_T$ distributions of $D^0$,
$D^+$, and $D^{*+}$ mesons measured at the Tevatron \cite{Acosta:2003ax}, as
may be seen in Fig.~\ref{fig:i}(a)--(c) and (d)--(f) on an absolute scale and
in the data-over-theory representation, respectively.

\section{Intrinsic charm from global analysis}
\label{sec:iii}

\begin{figure}[ht]
\begin{center}
\begin{tabular}{ccc}
\parbox{0.3\textwidth}{
\epsfig{file=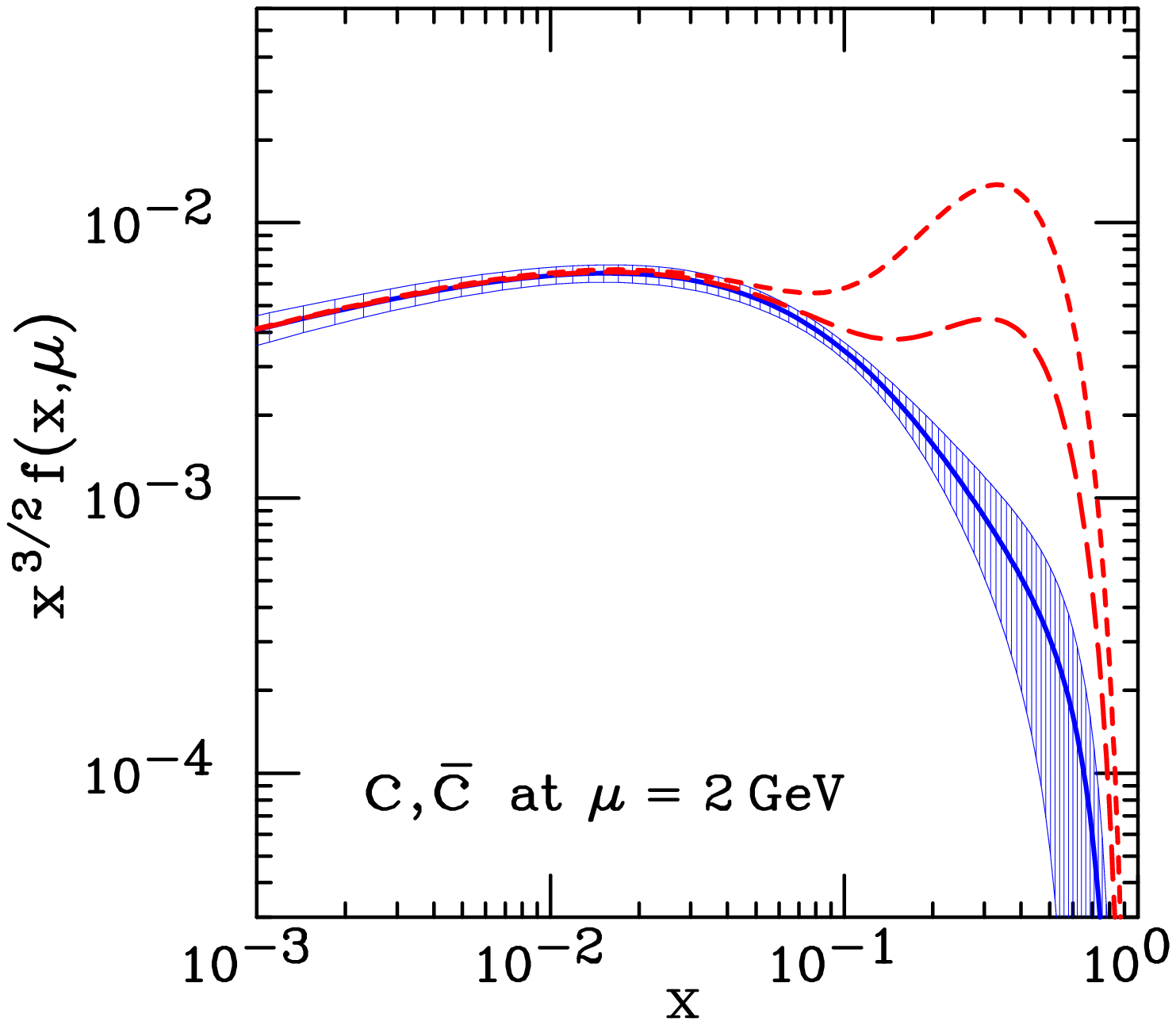,width=0.3\textwidth}}
&
\parbox{0.3\textwidth}{
\epsfig{file=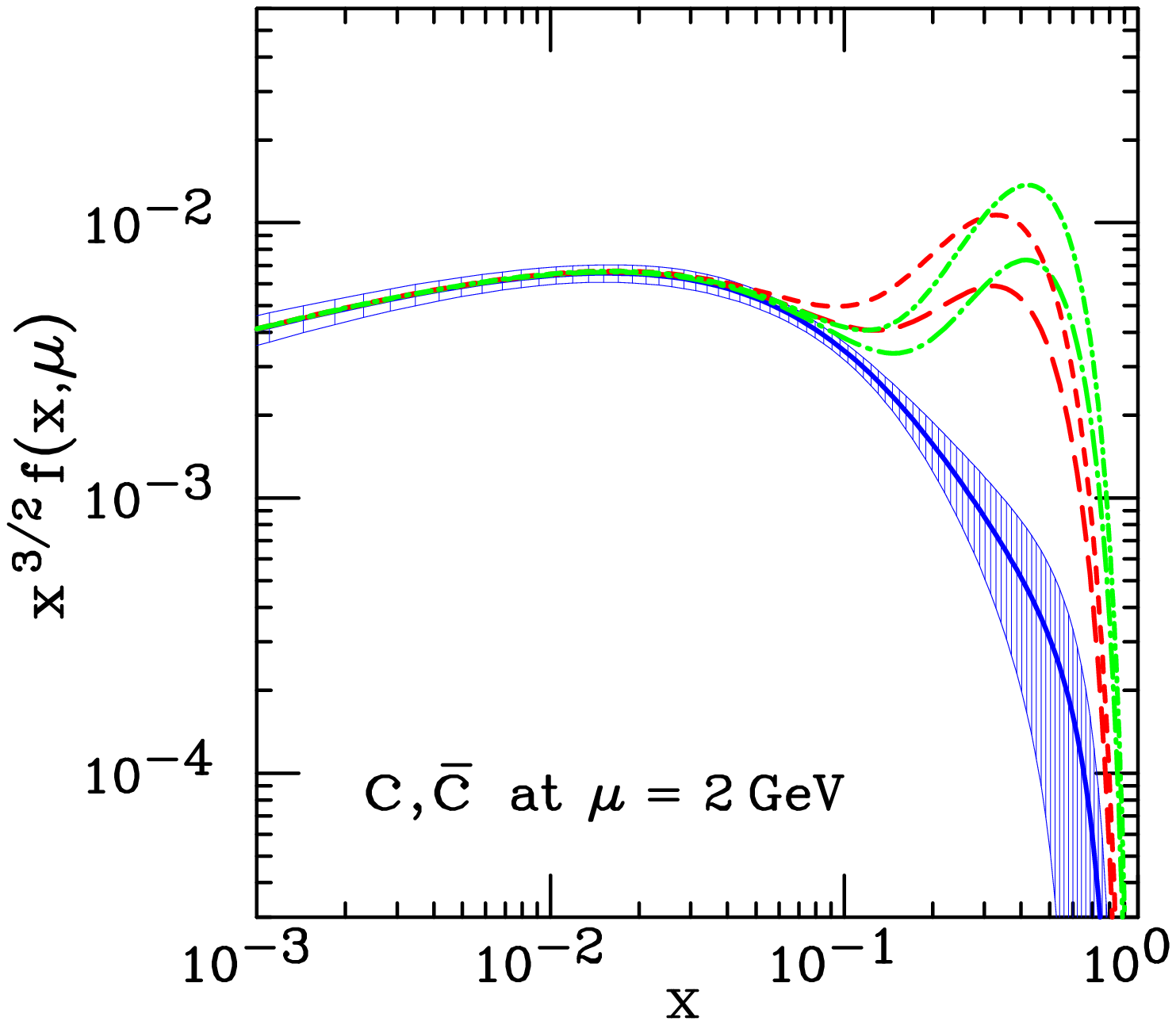,width=0.3\textwidth}}
&
\parbox{0.3\textwidth}{
\epsfig{file=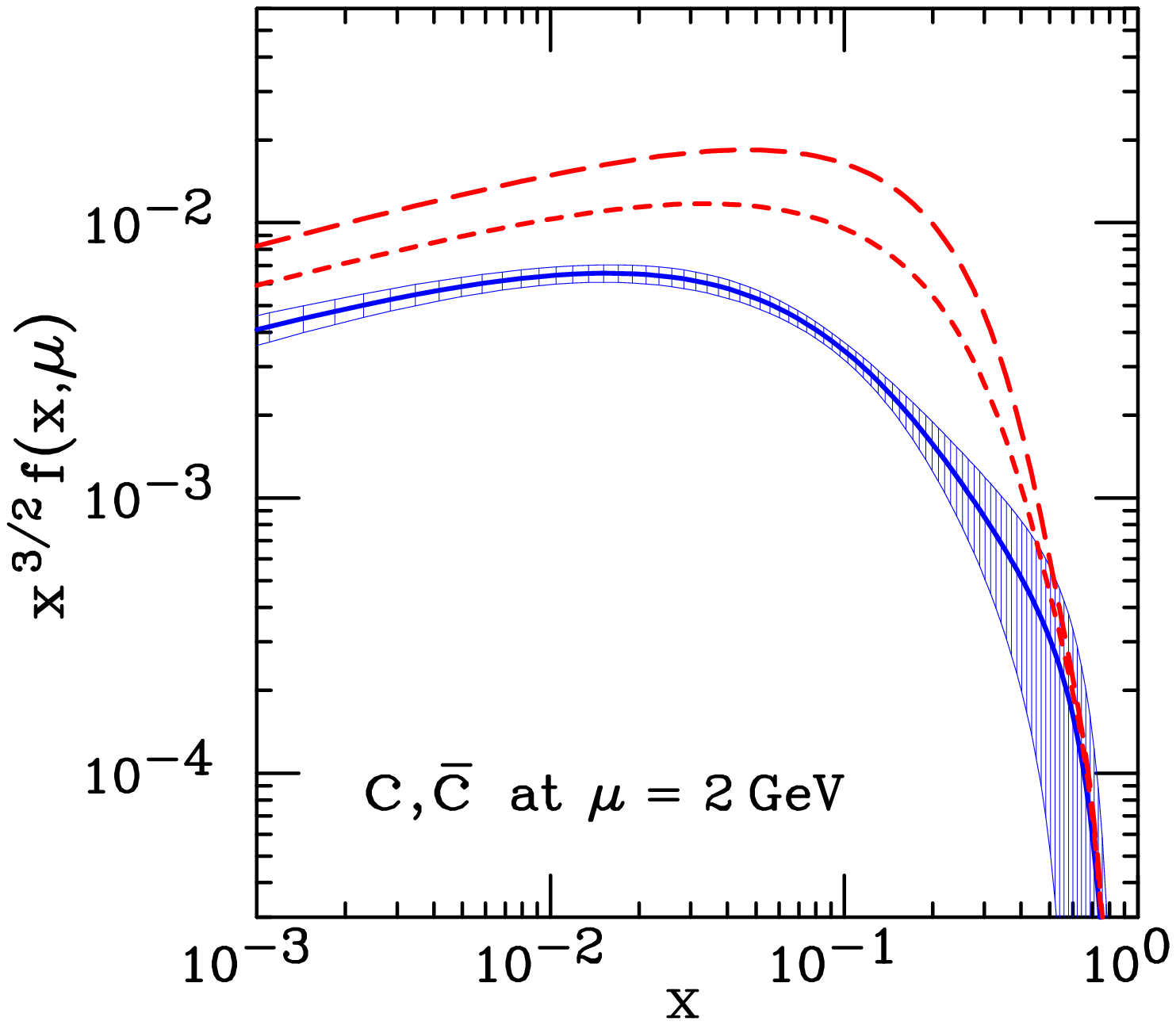,width=0.3\textwidth}}
\\
(a) & (b) & (c)
\end{tabular}
\end{center}
\caption{\label{fig:ii}%
The lower and upper dashed lines correspond to IC with modest and marginal
$\langle x\rangle_{c+\overline{c}}$ values, respectively, and the solid ones to
zero IC.}
\end{figure}

In Ref.~\cite{Pumplin:2007wg}, the CTEQ6.5 global analysis \cite{Tung:2006tb}
was repeated allowing for IC according to the
Brodsky-Hoyer-Peterson-Sakai (BHPS) \cite{Brodsky:1980pb},
meson-cloud \cite{Navarra:1995rq}, and
sea-like \cite{Pumplin:2007wg} models, with a modest and a marginal value of
$\langle x\rangle_{c+\overline{c}}$ in each case.
The $x$ distributions at $\mu=2$~GeV of the resulting $c$ and $\overline{c}$
parton distribution functions (PDFs) are shown in Fig.~\ref{fig:ii}(a)--(c)
for the BHPS, meson-cloud, and sea-like models, respectively. 

\section{Probing intrinsic charm at Tevatron and RHIC}
\label{sec:iv}

\begin{figure}[h!]
\begin{center}
\begin{tabular}{ll}
\parbox{0.45\textwidth}{
\epsfig{file=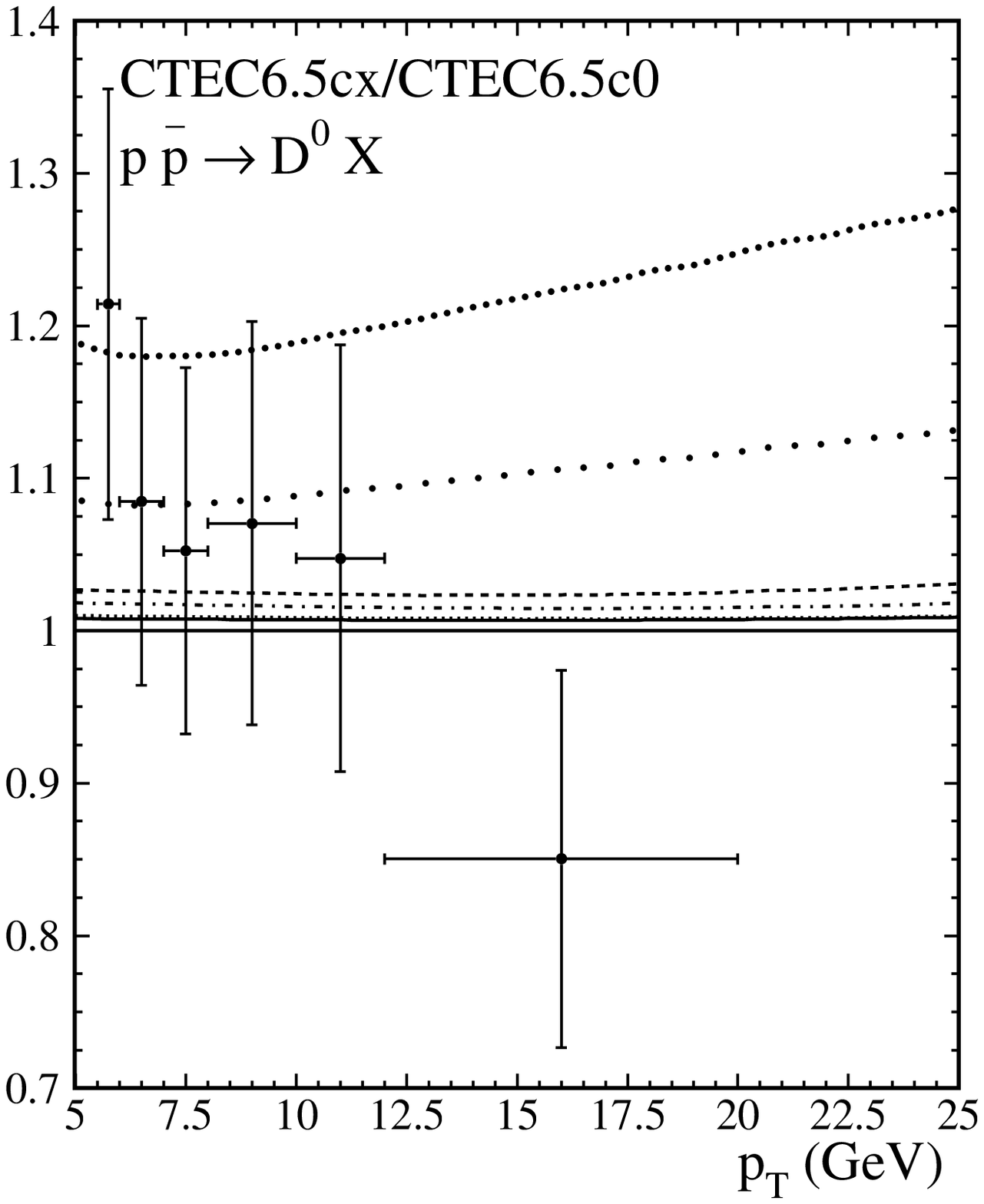,width=0.45\textwidth}}
&
\parbox{0.45\textwidth}{
\epsfig{file=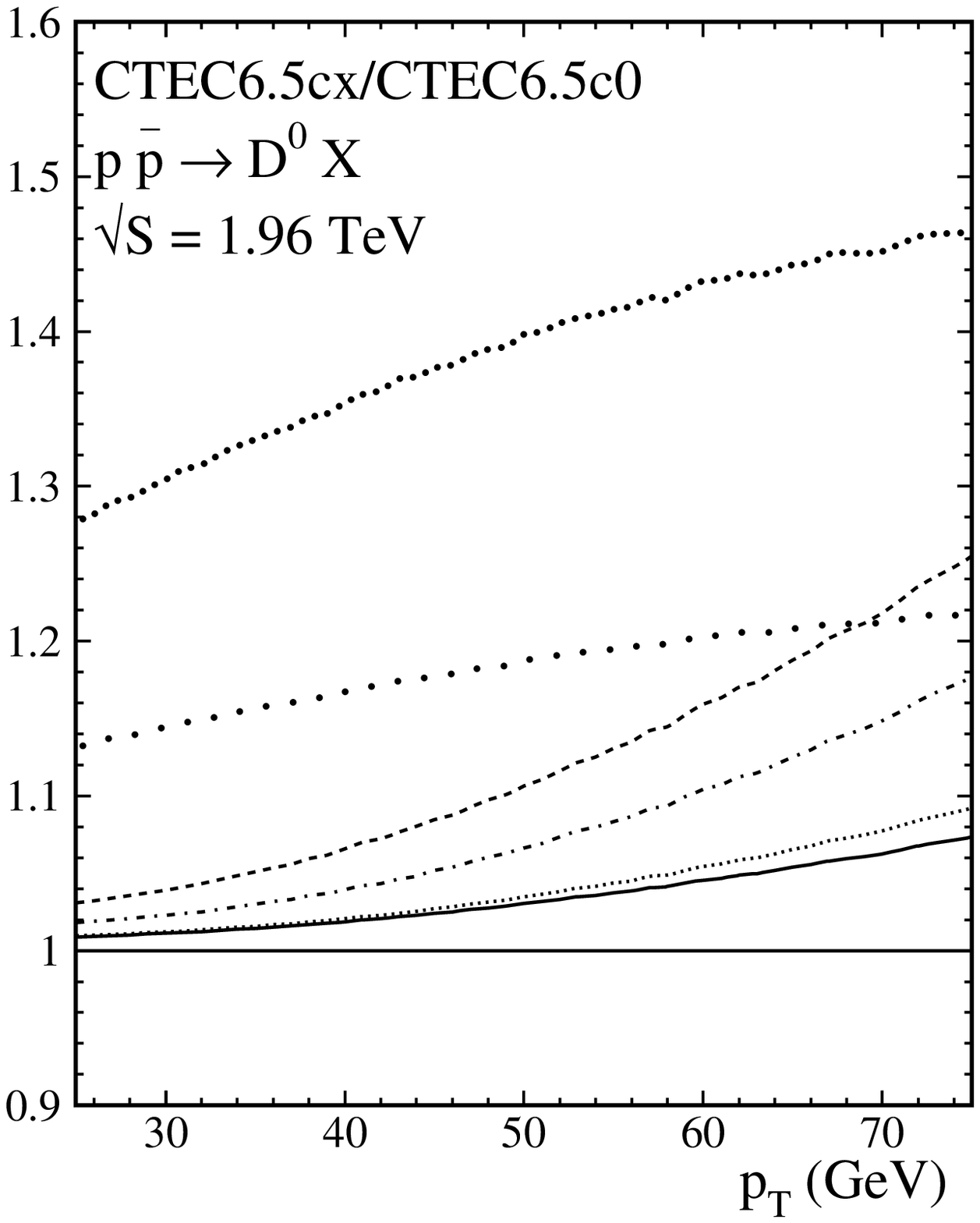,width=0.45\textwidth}}
\\
(a) & (b) \\
\parbox{0.45\textwidth}{
\epsfig{file=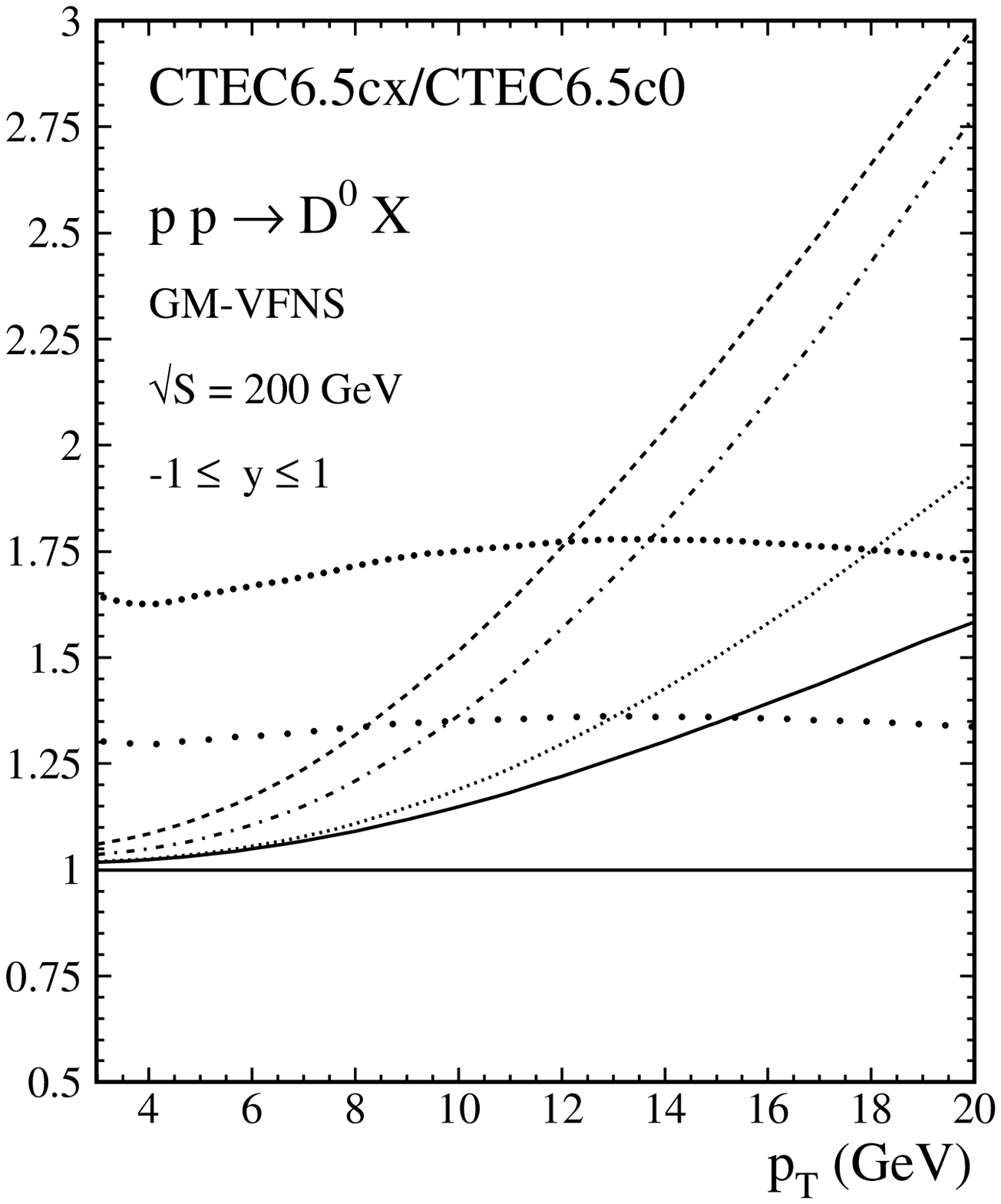,width=0.45\textwidth}}
&
\parbox{0.45\textwidth}{
\epsfig{file=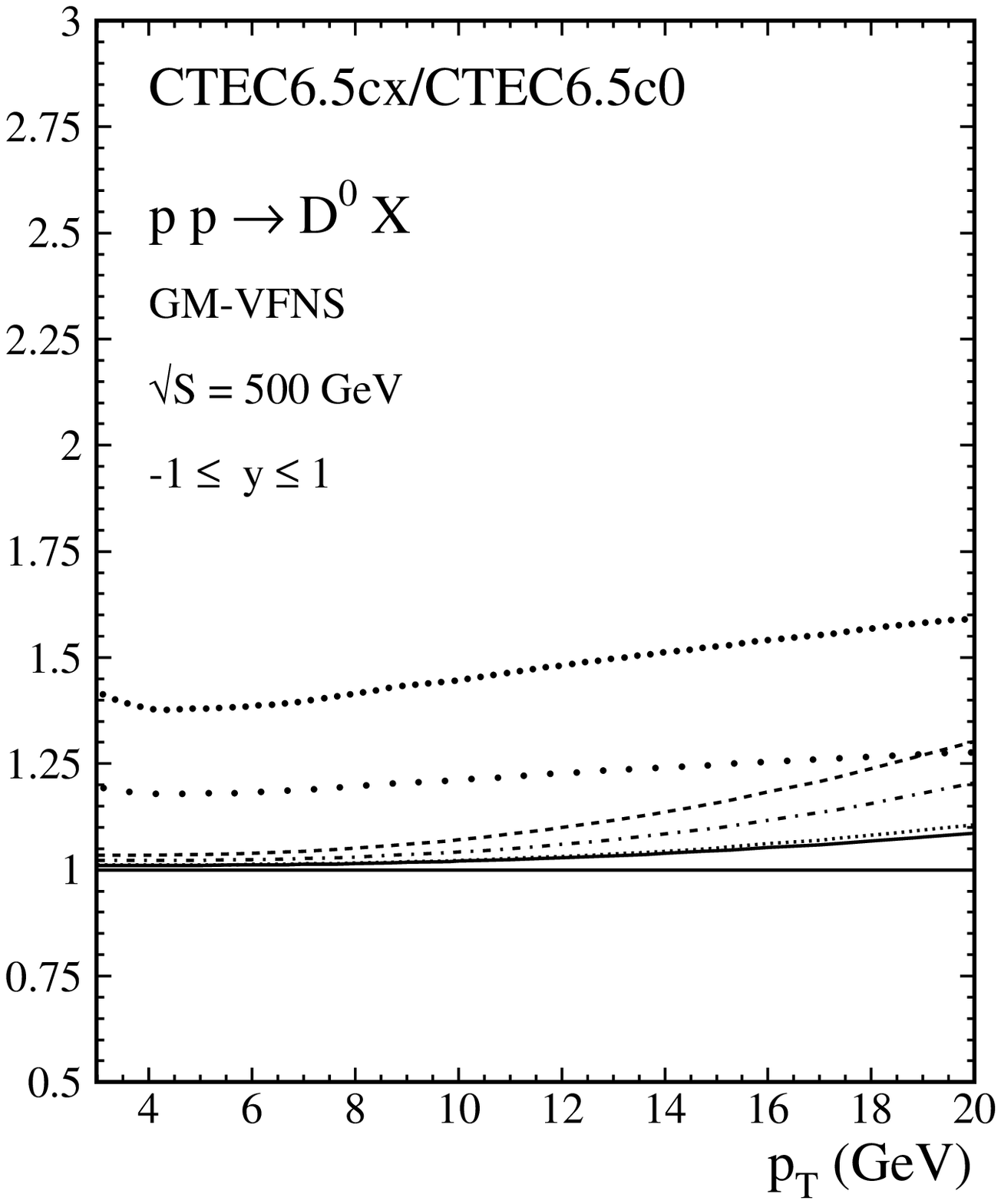,width=0.45\textwidth}}
\\
(c) & (d)
\end{tabular}
\end{center}
\caption{\label{fig:iii}%
The solid (dashed), densely dotted (dot-dashed), and scarcely dotted (dotted)
lines correspond to the BHPS \cite{Brodsky:1980pb}, meson-cloud
\cite{Navarra:1995rq}, and sea-like \cite{Pumplin:2007wg} models with modest
(marginal) values of $\langle x\rangle_{c+\overline{c}}$.}
\end{figure}

In the GM-VFNS, $D$-meson hadroproduction receives sizeable contributions from
partonic subprocesses with incoming $c$ or $\overline{c}$ quarks
\cite{Kniehl:2004fy,Kniehl:2005st} and is thus sensitive to their PDFs and
IC-induced enhancements in them.
In Fig.~\ref{fig:iii}(a), the relative enhancements due to the six IC
implementations of Ref.~\cite{Pumplin:2007wg} are compared with the $D^0$ data
of Ref.~\cite{Acosta:2003ax}.
We observe that the two large-$p_T$ data points tend to disfavour the sea-like
IC with marginal value of $\langle x\rangle_{c+\overline{c}}$, although a firm
statement would be premature.
On the other hand, the effects due the IC of the BHPS or meson-cloud models are
too feeble to be resolved by the presently published CDF data.

Since IC typically receives a large fraction $x$ of momentum from the parent
hadron, especially in the BHPS and meson-cloud models, and the kinematical
upper bound on $x$ scales with $x_T=2p_T/\sqrt{s}$, the sensitivity to IC may
be enhanced by increasing $p_T$ or lowering the cm.\ energy $\sqrt{s}$.
Keeping in mind that the CDF analysis of Ref.~\cite{Acosta:2003ax} is merely
based on 5.8~pb$^{-1}$ of data recorded in February and March 2002 and that,
since then, the integrated luminosity of run II has increased by more than a
factor of 1000, exceeding 6~fb$^{-1}$ as it does, significantly more precise
data at larger $p_T$ values are expected to come, with an increased sensitivity
to IC, as may be seen from Fig.~\ref{fig:iii}(b).
A hadron collider with smaller c.m.\ energy is in operation, namely RHIC with
presently $\sqrt s=200$~GeV, where the $p_T$ distribution of $pp\to X_c+X$
could be measured.
So far, only the STAR Collaboration has published $D$-meson production data,
namely for $d\mathrm{Au}\to D^0+X$ with $p_T<2.5$~GeV \cite{Adams:2004fc}.
In Fig.~\ref{fig:iii}(c), we observe significant enhancements due to IC in
$pp$ collisions at $\sqrt s=200$~GeV, by up to 75\% at $p_T=10$~GeV.
The high-energy $pp$ mode of RHIC, with $\sqrt{s}=500$~GeV, will accrue more
luminosity, perhaps 200--400~pb$^{-1}$.
However, this will happen at the expense of lowering the accessible $x$ values
and, thus, of reducing the relative IC effects on the cross sections,
especially for the BHPS and meson-cloud models, as is evident from
Fig.~\ref{fig:iii}(d).

\section{Conclusions}
\label{sec:v}

In the case of the Tevatron, we found that the BHPS and meson-cloud models
yield insignificant deviations from the zero-IC predictions, while the shift
produced by the sea-like model is comparable to the experimental error and
tends to worsen the agreement between theory and experiment.
However, the experimental errors in Ref.~\cite{Acosta:2003ax} are still too
sizeable to rule out this model as implemented in Ref.~\cite{Pumplin:2007wg}.
This is likely to change once the full data sample of run~II is exploited.

As for $pp$ collisions with $\sqrt{s}=200$~GeV at RHIC, all three IC models
predict sizeable enhancements, by up to 75\% at $p_T=10$~GeV.
In a future high-energy $pp$-collision mode of RHIC, with $\sqrt{s}=500$~GeV,
which is to accrue more luminosity, the cross sections would be increased,
while the relative shifts due to IC would be considerably smaller for the
BHPS and meson-cloud models.

\section*{Acknowledgments}

The author thanks G. Kramer, I. Schienbein, and H. Spiesberger for the
collaboration on the work presented here.
This work was supported in part
by the Bundesministerium f\"ur Bildung und Forschung BMBF through Grant No.\
05~HT6GUA,
by the Deutsche Forschungsgemeinschaft DFG through Grant No.\ KN~365/7--1, and
by the Helmholtz-Gemeinschaft Deutscher Forschungszentren HGF through Grant
No.~HA--101.
 

\begin{footnotesize}



%

\end{footnotesize}


\end{document}